\documentclass[superscriptaddress,preprint,secnumarabic,amssymb,nobibnotes,aps,prl]{revtex4}
\usepackage{graphicx}
\usepackage{bm}
\newcommand{\bea}{\begin{eqnarray}}
\newcommand{\eea}{\end{eqnarray}}
\newcommand{\be}{\begin{equation}}
\newcommand{\ee}{\end{equation}}
\newcommand{\bt}{\begin{tabular}}
\newcommand{\et}{\end{tabular}}

\newcommand{\no}{\nonumber}

\newcommand{\beas}{\begin{eqnarray*}}
\newcommand{\eeas}{\end{eqnarray*}}

\begin{document}
\date{\today}
\title{Impact of baryon resonances on the chiral phase transition at finite 
temperature and density}

%%%%%%%%%%%%%%%%%%%%%%%%%%%%%%%%%%%%%%%%%%%%%%%%%%%%%%%%%%%%%%%%%
\author{D.~Zschiesche} 
\affiliation{Institut f\"ur Theoretische Physik,
        Robert-Mayer-Str.\ 8-10, D-60054 Frankfurt am Main, Germany}

\author{G.~Zeeb}
\affiliation{Institut f\"ur Theoretische Physik,
        Robert-Mayer-Str.\ 8-10, D-60054 Frankfurt am Main, Germany}

\author{S.~Schramm}
\affiliation{Institut f\"ur Theoretische Physik,
        Robert-Mayer-Str.\ 8-10, D-60054 Frankfurt am Main, Germany}

\author{H.~St\"ocker}
\affiliation{Institut f\"ur Theoretische Physik,
        Robert-Mayer-Str.\ 8-10, D-60054 Frankfurt am Main, Germany}
\affiliation {Frankfurt Institute for Advanced Studies (FIAS),
Robert-Mayer-Str.\ 10, 
60054 Frankfurt am Main, Germany}

\begin{abstract}
We study the phase diagram of a
generalized chiral SU(3)-flavor model in mean-field approximation. 
In particular, the 
influence of the baryon resonances, and their couplings to 
the scalar and vector fields, 
on the characteristics of 
the chiral phase transition as a function of temperature and baryon-chemical
potential is investigated. Present and future finite-density lattice 
calculations might constrain the couplings of the fields to the baryons. 
The results are compared to recent lattice QCD calculations 
and it is shown that it is non-trivial to obtain, simultaneously, 
stable cold nuclear matter.
\end{abstract}

\maketitle

\section{I. Introduction}
QCD is the accepted underlying theory of strong interactions. 
It exhibits a very diverse and fascinating phase structure.
At high temperatures or chemical potentials two different 
significant changes 
in the structure of matter are expected: Deconfinement \cite{Svet82} 
and chiral
symmetry restoration \cite{namb61,Kirz72,Wein74}. 
Relativistic 
heavy ion collisions and astrophysical objects like neutron stars 
are two %The most
important experimental windows 
to gain information about these transitions,
where extreme conditions of temperature and/or density occur. 
In the hadronic world, as well as
in the region not too far above the phase transition 
to the deconfined and chirally restored 
phase, the strong coupling constant is large and thus QCD can not  
be treated perturbatively. 
From a theoretical point of view, 
lattice gauge calculations represent the most direct 
approach to investigate the QCD phase diagram of strongly interacting matter.
In the last years many lattice results on the phase 
structure of QCD at finite temperature $T$ 
and (recently) also at finite chemical potential $\mu$ were obtained
(see e.g.\  \cite{Fodor01,Karsch01,Fodor02,deForcrand02,Karsch03,Karsch03b,Karsch2004_a,KarschTawfika,KarschTawfikb,Fodor04}).
In particular, the properties of the 
chiral and deconfinement phase transitions and thermodynamic 
observables like pressure and energy density were investigated. 
However, lattice QCD alone does not seem to be able to completely disentangle
the physics of the QCD phase transition and give an explanation of 
the structure and the scales of the phase diagram. 
Uncertainties in the lattice results remain, 
as e.g.\ the large pion mass, 
large lattice spacings, or the reliability of the expansion schemes and 
the continuum limit \cite{deForcrand02}. 
Furthermore,   
in order to understand the dynamics of relativistic heavy ion collisions 
and the structure of neutron stars, the equation of state of 
strongly interacting matter is needed also at very high $\mu$.
%a region lattice QCD up to now and probably also in the near future 
%can not describe reliably.
This is a region lattice QCD can not describe reliably up to now 
and probably neither in the near future.
Thus, effective lagrangians \cite{Wein67,Cole69}  are studied, 
representing a complementary approach to 
disentangle and understand the 
physics of the QCD phase transition
(see e.g
\cite{Kirz72,Lee74,Baym77,Pisarski84,campbell90a,campbell90b,Ellis91,Brown91,Patkos91,Kusaka92,Ellis92,Kusaka93,Schaffner99,Scavenius00,Roder03,Dumitru03}
or Refs.\ in \cite{Stephanov04}). 

At high temperatures, so-called  
hadron resonance models represent 
a very successful effective approach.
The hadronic Bootstrap model \cite{Hagedorn65} 
predicted the existence of a limiting temperature for hadronic matter
long before lattice QCD provided first evidence for the 
transition to the deconfined, chirally restored phase \cite{Karsch01}.
Furthermore, for quark masses currently used in lattice calculations 
a resonance gas model with a 
percolation criterion gives a reasonable description of lattice data
 close to $T_c$ \cite{Karsch01}. 
As was shown in \cite{KarschTawfika,KarschTawfikb}, 
the density dependence of the QCD equation of 
state in the hadronic phase observed in recent lattice studies
can be understood in terms of a baryonic resonance gas.
Finally, the critical temperature $T_c (\mu=0)$ 
from lattice QCD depends only weakly on changes of the 
lightest hadron masses 
\cite{Karsch01},
in contrast to the predictions of a linear $SU(2) \times SU(2)$ 
linear $\sigma$ model \cite{Dumitru03} without resonances. 
The pressure of heavy states, however, may reduce the dependence of 
$T_c$ on the quark masses \cite{Gerber89}, in accord with the findings from 
the lattice. Furthermore, we point out that typically models relying mainly 
on order-parameter (infra-red) dynamics and which do not include more massive 
states predict significantly smaller phase transition temperatures in 
baryon-dense matter than obtained on the lattice 
(see e.g.\ Fig.\ 6 in \cite{Stephanov04}).

These findings indicate  
that a hadron resonance gas approach is reasonable below $T_c$ and that 
in such an approach the contributions of heavy resonances are very 
important.
Even though the free resonance gas gives a very good description of the 
lattice data -- the same lattice data unambigiously show that there are 
temperature and medium effects, for example the change in the chiral 
condensate \cite{Karsch01}. This is not accounted for 
in non-interacting approaches. In addition, 
studies of the nucleon-nucleon interaction and 
dense nuclear matter in boson exchange models 
show the importance 
of various meson exchanges \cite{Mach89}. 
Many properties of finite nuclei 
and of nuclear matter saturation
can be understood in terms of scalar and vector potentials 
\cite{Duer56,Clar73,Wale74,Ana83}.

Furthermore, it is desirable that %the
effective models
incorporate %the 
some known features of QCD.
Thus, an interacting hadron gas accounting for chiral symmetry 
restoration and other known medium effects should be investigated,
like it has been done e.g.\ in \cite{Ellis91}.
%is considered. 
To this end, we consider here the phase transition properties of
the chiral hadronic model presented in \cite{paper3}.
It represents a relativistic field theoretical model of baryons and
mesons built on chiral $SU(3)_L \times SU(3)_R$
symmetry and broken scale invariance.
A non-linear realization (see \cite{Wein68,ccwz})
of chiral symmetry
is adopted. 
The model has been shown to 
successfully describe hadronic vacuum properties, nuclear matter saturation, 
finite nuclei and hypernuclei 
\cite{paper3,Schramm02,Beckmann01}. Furthermore, it has been applied to
the description of hot and cold non-strange and strange 
hadronic matter \cite{Zschiesche:2000ew}, the structure of 
rotating neutron stars \cite{Schramm02b,Schramm02c} 
and observables in relativistic heavy ion collisions
\cite{Zschiesche02a,zsch01}.
In Refs.\ \cite{panic99,zsch01} it was shown that
%at high densities or temperatures
a chiral phase transition to a chirally restored 
phase occurs
at high densities or temperatures. %G.
At vanishing chemical potential the critical temperature is
%in the range between
$150-180$ MeV, i.e.\ in the range 
predicted by lattice calculations.
The nature of this transition -- at finite temperature as well as at finite 
chemical potential -- crucially depends on the 
number of degrees of freedom coupled to the mesonic fields and 
on the strength of these couplings \cite{Thei83,Wald87,zsch01}. 
For a deeper understanding 
of the chiral phase transition in strongly interacting 
matter, we will thus systematically analyze the role of 
baryon resonances, which 
in our investigation are
effectively represented 
by the lightest baryon decuplet.

Note that only the {\em chiral}\/ transition is adressed by our model 
since only hadronic degrees of freedom are considered. %G.
%Note that since only hadronic degrees of freedom are considered in 
%our approach, only the chiral transition may be investigated.
Nevertheless, we can ask when particle densities 
are that large  
that hadrons should no longer be 
the suitable degrees of freedom anymore. 
It turns out that 
the coupling of the baryon decuplet may give the right scale for 
the critical temperature and also 
lead to a drastically increasing energy- and baryon density at the 
phase boundary.
By simultaneously studying
different observables, like here the predicted phase diagram and 
nuclear matter saturation, it is possible to relate these different 
aspects %states
 of strongly interacting matter. 
This should on one hand help to get a deeper understanding  
of the different regions of the QCD phase diagram 
and on the other hand provide more constraints on    
effective models, especially in the resonance sector but also, 
for example, on the form of the potentials. 
Furthermore, we can study how the restoration 
and the different symmetry breaking patterns are reflected in 
the properties of the model under different external conditions.

The paper is organized as follows. In section II we introduce the chiral 
SU(3) model. Section III shows the results for the phase
diagram and for thermodynamic observables. In section IV we conclude and
give an outlook to future work.

\section{II. The chiral model}

The chiral hadronic SU(3) lagrangian in the mean field approximation 
has the following basic structure
\begin{equation}
{\cal L} = {\cal L}_{\rm kin} + {\cal L}_{\rm BM} + {\cal L}_{\rm
BV} + {\cal L}_{\rm vec} + {\cal L}_0 + {\cal L}_{\rm SB}  ~,
\end{equation}
consisting of interaction terms between baryons respectively
spin-0 (BM) and spin-1 (BV) mesons
\begin{eqnarray}
{\cal L}_{\rm BM}+{\cal L}_{\rm BV} &=& -\sum_{i}
\overline{\psi}_i \left[ g_{i\sigma} \sigma + g_{i\zeta} \zeta + g_{i \omega}\gamma_0 \omega^0 +
g_{i \phi}\gamma_0 \phi^0
\right] \psi_i~, \nonumber \\
{\cal L}_{\rm vec} &=& \frac{1}{2}
m_{\omega}^2\frac{\chi^2}{\chi_0^2} \omega^2 + \frac{1}{2}
m_\phi^2\frac{\chi^2}{\chi_0^2} \phi^2 
+ g_4^4 (\omega^4 + 2 \phi^4 )
\end{eqnarray}
summing over the baryonic octet (N,$\Lambda$,$\Sigma$,$\Xi$), and
decuplet ($\Delta$,$\Sigma^\ast$,$\Xi^\ast$,$\Omega$). 
The interactions between the scalar mesons 
(with the scale breaking terms containing the dilaton field $\chi$) %G. 
read 
\begin{eqnarray}
{\cal L}_0 &=& -\frac{1}{2} k_0 \chi^2 (\sigma^2+\zeta^2) + k_1
(\sigma^2+\zeta^2)^2 + k_2 ( \frac{ \sigma^4}{2} + \zeta^4)
+ k_3 \chi \sigma^2 \zeta \nonumber \\
& & {}- k_4 \chi^4 - \frac{1}{4}\chi^4 \ln \frac{ \chi^4 }{
\chi_0^4}
+\frac{\delta}{3}\ln \frac{\sigma^2\zeta}{\sigma_0^2 \zeta_0}~.
\end{eqnarray}
An explicit symmetry breaking term mimics the QCD effect of non-zero 
current quark masses
\begin{eqnarray}
{\cal L}_{\rm SB} &=& -\left(\frac{\chi}{\chi_0}\right)^2
\left[m_\pi^2 f_\pi \sigma + (\sqrt{2}m_K^2 f_K -
\frac{1}{\sqrt{2}}
m_{\pi}^2 f_{\pi})\zeta \right]~.
\end{eqnarray}
The term ${\cal L}_{\rm kin}$ in (1) contains the kinetic energy terms of the hadrons.
The general model incorporates the full
lowest baryon (octet and decuplet) and meson multiplets. 
Here, instead, we only consider the mesons relevant for symmetric nuclear
matter, namely the scalar field $\sigma$ and its $s\bar{s}$ counterpart $\zeta$
(which can be identified with the observed $f_0$ particle), as well as
the $\omega$ and $\phi$ vector mesons.
All other mesons as well as
heavier baryon resonances are treated as free particles and thus do
not act as sources of the field equations.
The term ${\cal L}_{\rm vec}$ generates the masses of the spin-1 mesons through
the interactions with spin-0 mesons. The scalar interactions ${\cal L}_0$
induce the spontaneous chiral symmetry breaking. Another scalar %,
isoscalar field,
the dilaton $\chi$, which simulates the breaking of the QCD scale invariance,
can be identified with the gluon condensate \cite{sche80} (for a more
detailed discussion see \cite{paper3}).
The effective masses
$m_i^*(\sigma,\zeta)=g_{i\sigma}\,\sigma+g_{i\zeta}\,\zeta$
of the baryons are generated through their coupling to the scalar
fields, which attain non-zero vacuum expectation values due to the
self-interactions \cite{paper3}
\footnote{For example in \cite{Ellis91} 
the quark-hadron phase transition was 
investigated in an effective lagrangian approach only containing 
broken scale invariance. In this work we keep the gluon condensate
frozen and thus focus on the role of the quark condensates. In a
future work the case with a dynamical dilaton field will be addressed.}. 
For the decuplet $D$ we introduce 
an explicitly symmetry breaking mass term of the form
\be
{\cal L} = - m_{\rm Dec} \bar{D} D \,.
\ee
With this term, it is possible to systematically study 
how the strength of the scalar coupling of the resonances influences 
the properties of hadronic matter, as will be discussed 
in more detail. % below.
%In addition, 
In the strange sector 
an additional %G
symmetry breaking term of the form 
$m_3 (\sqrt{2} (\sigma-\sigma_0) + (\zeta-\zeta_0))$
is introduced, %G
where %with
$m_3=1.25$. %is introduced, 
It is chosen in accord with the octet sector \cite{paper3,zsch01}, and %which 
guarantees that the decuplet masses always stay above 
the corresponding octet masses
\footnote{To discuss the phase diagram, different symmetry breaking terms
in the strange sector would have been possible.
However, by choosing the same term as in former
work we could make contact to the parameter studies made
there. The difference to choosing an explicit symmetry
breaking of the form $m_3 n_s (\sqrt{2} (\sigma-\sigma_0) + (\zeta-\zeta_0))$
is very small.}.
%The two extreme cases 
%$m_{\rm Dec}=0$ and $m_{\rm Dec}=1232 \mbox{ MeV}$ correspond 
%to the cases CII and CI in \cite{zsch01}, respectively
%\footnote{As already shown in \cite{zsch01}, the choice of this term 
%also influences the phase structure of the model. In particular,   
%two distinct phase transitions -- corresponding to seperate
%discontinuities in the non-strange and the strange condensate -- may
%occur. This will be discussed in detail in  a future work, but here we will
%focus on the general role of the baryon resonances and thus only
%consider one way of explicit symmetry breaking.}.

%In the strange sector of the baryon decuplet 
The resulting mass terms read 
\begin{eqnarray}
\label{resmassen}
 m_{\Delta}
   &=&  m_{\rm Dec} +
        g_D^S \left[(3-\alpha_{DS})\sigma+\alpha_{DS} \sqrt{2}\zeta\right] \no\\
 m_{\Sigma^\ast}
   &=& m_{\rm Dec} + m_3 (\sqrt{2}(\sigma- \sigma_0) + \zeta-\zeta_0) +
       g_D^S \left[ 2\sigma+ \sqrt{2}\zeta\right] \no\\
 m_{\Xi^\ast}
   &=& m_{\rm Dec} + m_3 (\sqrt{2}(\sigma- \sigma_0) + \zeta-\zeta_0) +
       g_D^S \left[(1+\alpha_{DS})\sigma+(2-\alpha_{DS})\sqrt{2}\zeta\right] \no\\
 m_{\Omega}
   &=& m_{\rm Dec} + m_3 (\sqrt{2}(\sigma-\sigma_0) + \zeta-\zeta_0) +
       g_D^S \left[2 \alpha_{DS}\sigma+(3-\alpha_{DS})\sqrt{2}\zeta\right] ~.
\end{eqnarray}
The two extreme cases 
$m_{\rm Dec}=0$ and $m_{\rm Dec}=1232 \mbox{ MeV}$ correspond 
to the parameter studies CII and CI in Ref.\ \cite{zsch01}, respectively
\footnote{As already shown in \protect{\cite{zsch01}}, 
the strength of the 
$m_3$-coupling  
also influences the phase structure of the model. In particular,   
two distinct phase transitions -- corresponding to seperate
discontinuities in the non-strange and the strange condensate -- may
occur. This will be discussed in detail in  a future work, but here we will
focus on the general role of the baryon resonances and thus only
consider one way of explicit symmetry breaking.}.

For a given value of the explicit symmetry breaking $m_{\rm Dec}$, the 
two couplings $g_D^S$ and $\alpha_{DS}$ are adjusted to the vacuum
masses of the decuplet resonances.
(Note that the %by G.
coupling of the $\Delta$ to the strange condensate is non-zero but small.)

The vector couplings $g_{i\omega}$ and $g_{i\phi}$ 
for the octet as well as for the decuplet result from pure $f$-type
coupling as discussed in
\cite{paper3,springer}, 
\begin{eqnarray}
 g_{i\omega} &=& (n^i_q-n^i_{\bar{q}}) g_{8,10}^V \nonumber\\
 g_{i\phi}   &=& -(n^i_s-n^i_{\bar{s}}) \sqrt{2} g_{8,10}^V\, ,
\end{eqnarray}
with $i=N,\Lambda,\Sigma,\Xi,\Delta, \Sigma^\ast, \Xi^\ast, \Omega$\,,
while %and
$g_8^V$ and $g_{10}^V$ denote the vector coupling of the baryon
octet and decuplet, respectively. 
$n^i$ represents the number of constituent quarks of 
a particular species in a given hadron, where the index $q$ 
represents the light 
$u$- and $d$-quarks, $s$ the strange quark, and $\bar{q}, \bar{s}$ the 
corresponding antiquarks.
The resulting relative couplings correspond to the additive quark model
constraints.

The grand canonical thermodynamic potential of the system
can be written as
\begin{equation}
\label{thermpot}
   \frac{\Omega}{V}= -{\cal L}_{\rm vec} - {\cal L}_0 - {\cal L}_{\rm SB}
-{\cal V}_{\rm vac} \\
\mp  T \sum_i \frac{\gamma_i }{(2 \pi)^3}
\int d^3k \left[\ln{\left(1 \pm e^{-\frac 1T[E^{\ast}_i(k)-\mu^{\ast}_i]}\right)}
\right],
\end{equation}
where $\gamma_i$ denote the 
hadronic spin-isospin degeneracy factors and 
$E^{\ast}_i (k) = \sqrt{k_i^2+{m_i^*}^2}$ are the
single particle energies.
The effective chemical potentials 
read
 $\mu^{\ast}_i = \mu_i-g_{i \omega} \omega-g_{i \phi} \phi$, 
with $\mu_i= (n^i_q - n^i_{\bar{q}}) \mu_q + (n^i_s - n^i_{\bar{s}}) \mu_s$. 
The vacuum energy ${\cal V}_{\rm vac}$ (the potential at $\rho_B=T=0$)
has been subtracted in 
order to get a vanishing vacuum energy.

By extremizing $\Omega/V$ one obtains self-consistent equations for the meson
fields. We here consider non-strange matter, i.e.,  
for given $T$- and $\mu_q$-values the strange chemical potential 
$\mu_s$ is chosen such that the net number of strange quarks 
in the system is zero. Then the dominant fields are 
the $\sigma$ and the $\omega$. Their field equations read
\begin{eqnarray}
\frac{\partial ({\cal L}_0)}{\partial \sigma} +
\frac{\partial ({\cal L}_{\rm SB})}{\partial \sigma} &=&
 \sum_i \frac{\partial m_i^{\ast}}{\partial \sigma}
\frac{\gamma_i}{(2\pi)^3}  \int d^3 k \frac {m_i^\ast}{E_i^\ast}
          \left( n_{k,i} + \bar{n}_{k,i} \right) 
\equiv \sum_i g_{i \sigma} \rho^s_i
\nonumber\\
\frac \chi\chi_0 m_{\omega}^2 \omega +  4 g_4^4 \omega^3 &=&
 \sum_i g_{i \omega} \frac{\gamma_i}{(2\pi)^3}  
\int d^3 k \left( n_{k,i} - \bar{n}_{k,i} \right) 
\equiv \sum_i g_{i \omega} \rho_i ,
\end{eqnarray}
with the fermionic (anti-)particle distribution functions 
\bea
n_{k,i} \equiv n_{k,i} (T,\mu_i^{\ast}) &=&
\frac {1} {e^{\frac{1}{T}[E_i^{\ast} - \mu_i^{\ast}]}+ 1} \\
\bar{n}_{k,i} \equiv \bar{n}_{k,i} (T,\mu_i^{\ast}) &=&
\frac {1} {e^{\frac{1}{T}[E_i^{\ast} + \mu_i^{\ast}]}+ 1} , 
\eea
while $\rho_i^s$ and $\rho_i$ denote the 
scalar and vector density of particle species $i$, respectively.

The sources for the scalar and vector fields mainly depend on two 
parameters: the number of degrees of freedom coupled to the fields 
and the corresponding coupling constants. For the octet, the
scalar couplings are fixed by the vacuum masses and the 
vector coupling is adjusted to give the correct 
%nuclear matter binding energy. 
binding energy of nuclear matter \cite{paper3}.

In the current investigation we will systematically study the
dependence of the hadronic matter properties on the strength
of the scalar and vector coupling for the decuplet.
These are controlled
%%For the decuplet we will in the current investigation systematically
%study the dependence of the hadronic matter properties on the strength
%of the scalar and vector coupling, controlled 
by the explicit symmetry breaking $m_{\rm Dec}$ and 
the relative strength of the vector coupling,  
\begin{equation}
\label{rv}
r_{\rm v}= \frac{g_{10}^V}{g_8^V}=\frac{g_{\Delta\omega}}{g_{N\omega}} , 
\end{equation}
respectively \footnote{The decuplet couplings are partially constrained by the 
condition that nuclear matter and not the chirally restored phase is stable
\cite{Wald87,koso98,Zschiesche:2000ew}.
However, since we want to focus here on the general influence of 
the baryon resonances on the phase transition, they are treated as free
parameters but it is always stated whether nuclear matter is
stable or not.}.
%\begin{eqnarray*}
%mR_i &=& g_{i\sigma} \sigma + g_{i\zeta} \zeta + m_{\rm Dec} \\
%r_{\rm v} &=& \frac{g_{\Delta\omega}}{g_{N\omega}}
%\end{eqnarray*}
The two parameters $m_{\rm Dec}$ and $r_{\rm v}$ 
determine the abundance and the contribution of the baryon resonances 
to the scalar and vector field equations and, as will be shown later,  
also determine the phase diagram of the model.
From Eq.\ (\ref{resmassen}) one sees that 
the larger the 
explicit symmetry breaking %ESB 
value $m_{\rm Dec}$, the smaller are the 
resulting decuplet-scalar couplings  $g_{i\sigma}\,,\ g_{i\zeta}$, 
i.e.\ the contribution of the resonances to the source terms of 
the field equations. 
In addition, a large explicit symmetry breaking $m_{\rm Dec}$ also gives 
a higher effective mass in the medium. 
Equation (\ref{rv}) shows that 
the larger $r_{\rm v}$, the larger 
is $g_{10}^V$, which leads to smaller effective potentials
$\mu_i^\ast$ for the decuplet states.
This reduces the net number of baryon resonances at a 
given chemical potential. 

\section{III. Results}

\begin{figure}[h]
\vspace*{-0.8cm}
\begin{center}
\centerline{\parbox[b]{8.5cm}{
\includegraphics[width=8.5cm]{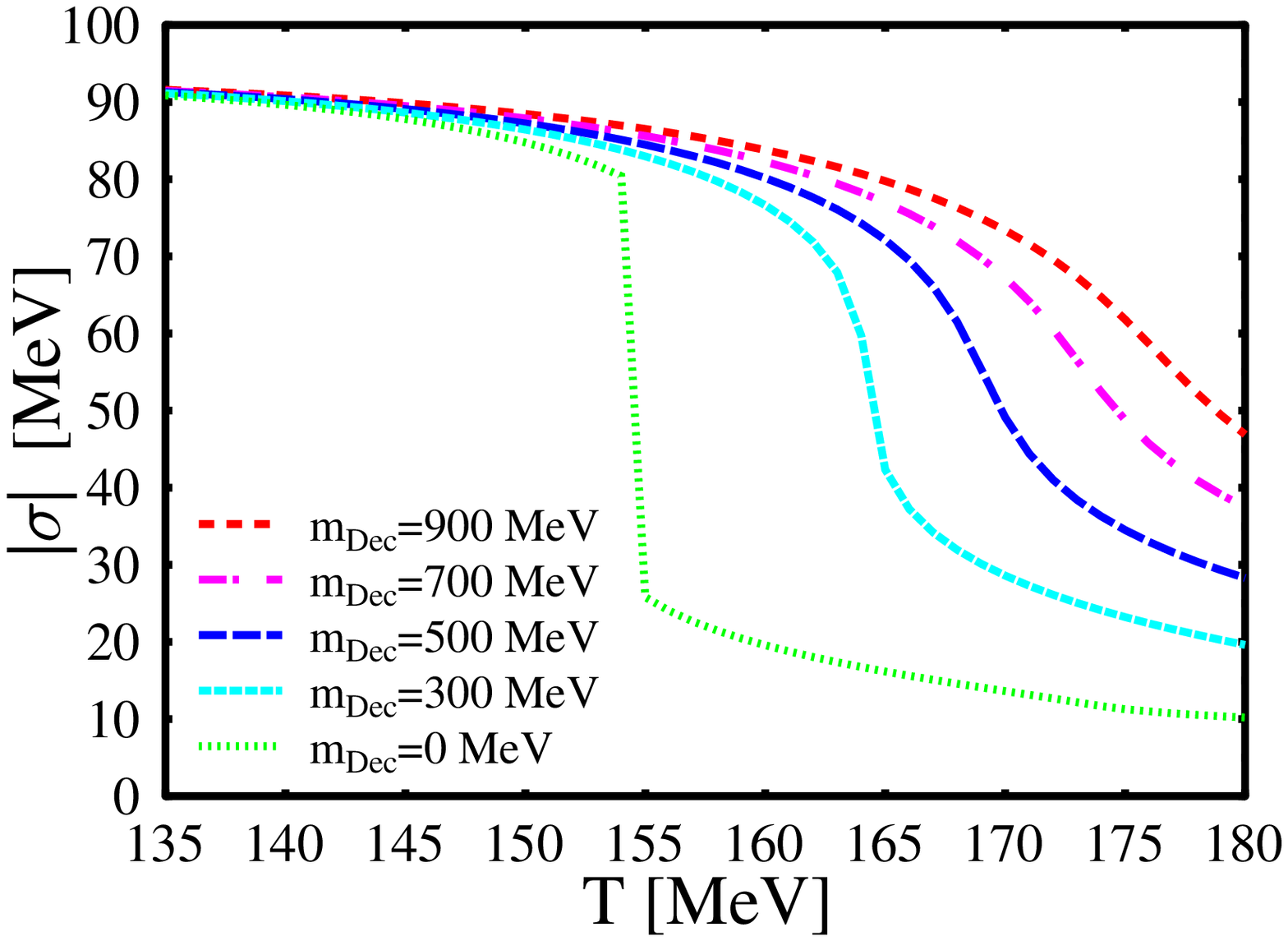}}
\parbox[b]{8.5cm}{
\includegraphics[width=8.5cm]{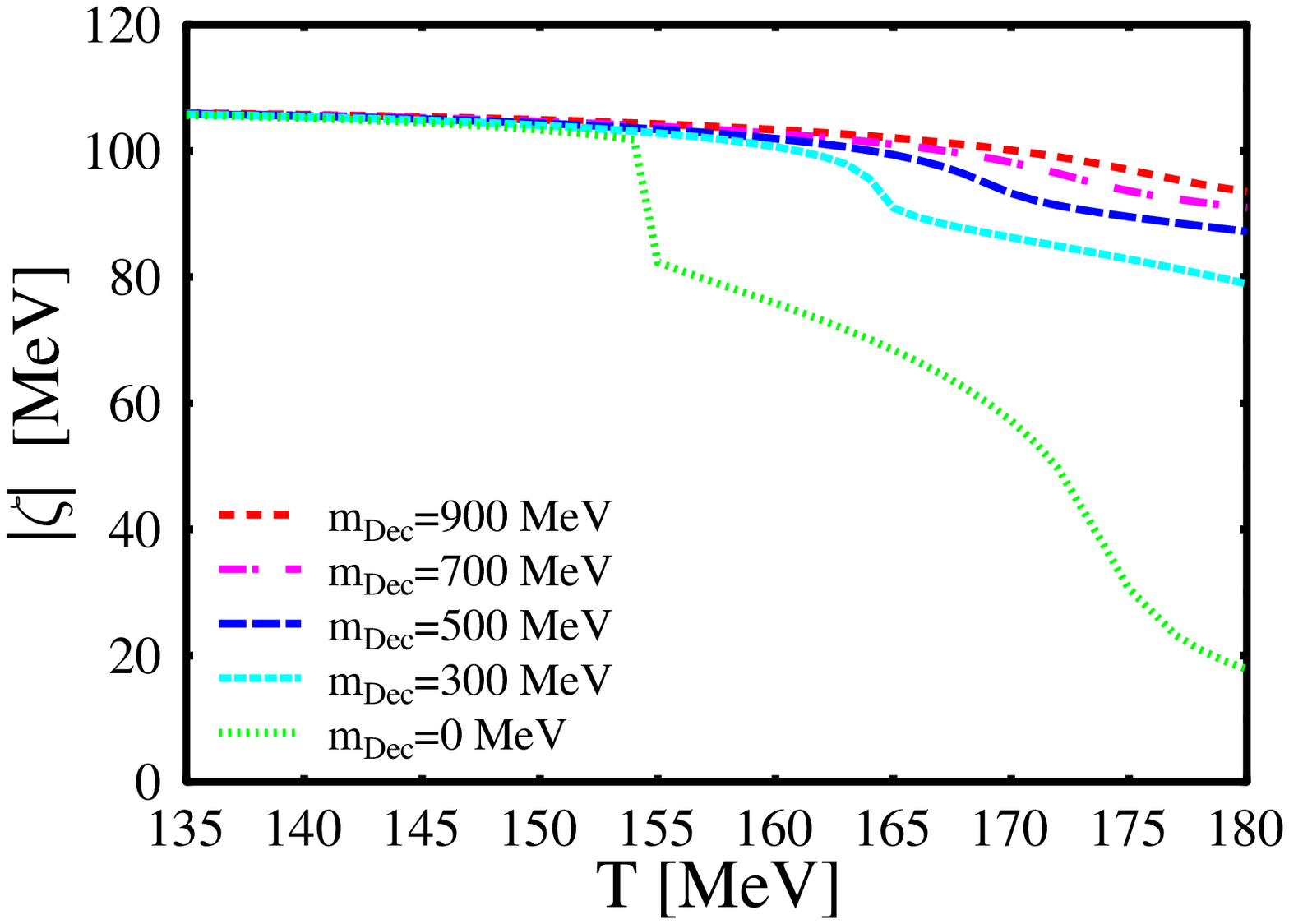}}
}
\vspace{-0.5cm}
\caption{Non-strange condensate $\sigma$ (left) and strange
condensate $\zeta$ (right) as a function of temperature for vanishing
chemical potential ($\mu_q$=0).
The curves correspond to different values of the explicit 
symmetry breaking term $m_{\rm Dec}$.
\label{mu0}}
\end{center}
\vspace{-0.5cm}
\end{figure}

First, we concentrate on the phase transition behavior at vanishing
chemical potential. Since the vector field $\omega$ 
couples to the vector densities $\rho_i$, it vanishes 
independently of the coupling chosen. Thus, the phase transition
behavior at $\mu=0$ 
depends only on the scalar coupling, i.e.\ on the choice of $m_{\rm Dec}$, 
not on $r_{\rm v}$. %G.
Figure \ref{mu0} shows the scalar fields $\sigma$ and $\zeta$ 
 as a function of 
temperature $T$ for different values of the scalar coupling. 
For $m_{\rm Dec} < 260$ MeV %,
(i.e.\ $g_{\Delta\sigma} \stackrel{>}{\sim} g_{N\sigma}$),
a first-order phase transition occurs, as 
indicated by the jump in the chiral condensates
at $T_c \approx 155$ MeV. %G.
For smaller couplings of the baryon resonances the scalar fields
decrease continuously and a crossover is observed.
The strange condensate $\zeta$ is much less affected around $T_c$ 
than the non-strange condensate $\sigma$, reflecting the fact that most
of the produced pairs are nucleons and deltas.
If the value of the additional symmetry breaking for the strange resonances
($m_3$ from Eq.\ \ref{resmassen}) %G.
is decreased -- re-adjusting the corresponding coupling strength to
keep the masses at their vaccuum expectation values --
these decuplet states contribute stronger and might even produce 
another %G.
first-order phase transition characterized by a discontinuity
in the strange condensate. This will be discussed in detail in 
\cite{zeeb04}.
%%%%%%%%%%%%%%%%%%%%%%%% condensates at mu=0

\begin{figure}[h]
%\vspace*{-3cm}
%\epsfysize=0.8\textheight
\vspace*{-0.8cm}
\begin{center}
\centerline{\parbox[b]{8.5cm}{
\includegraphics[width=9.2cm,height=6cm]{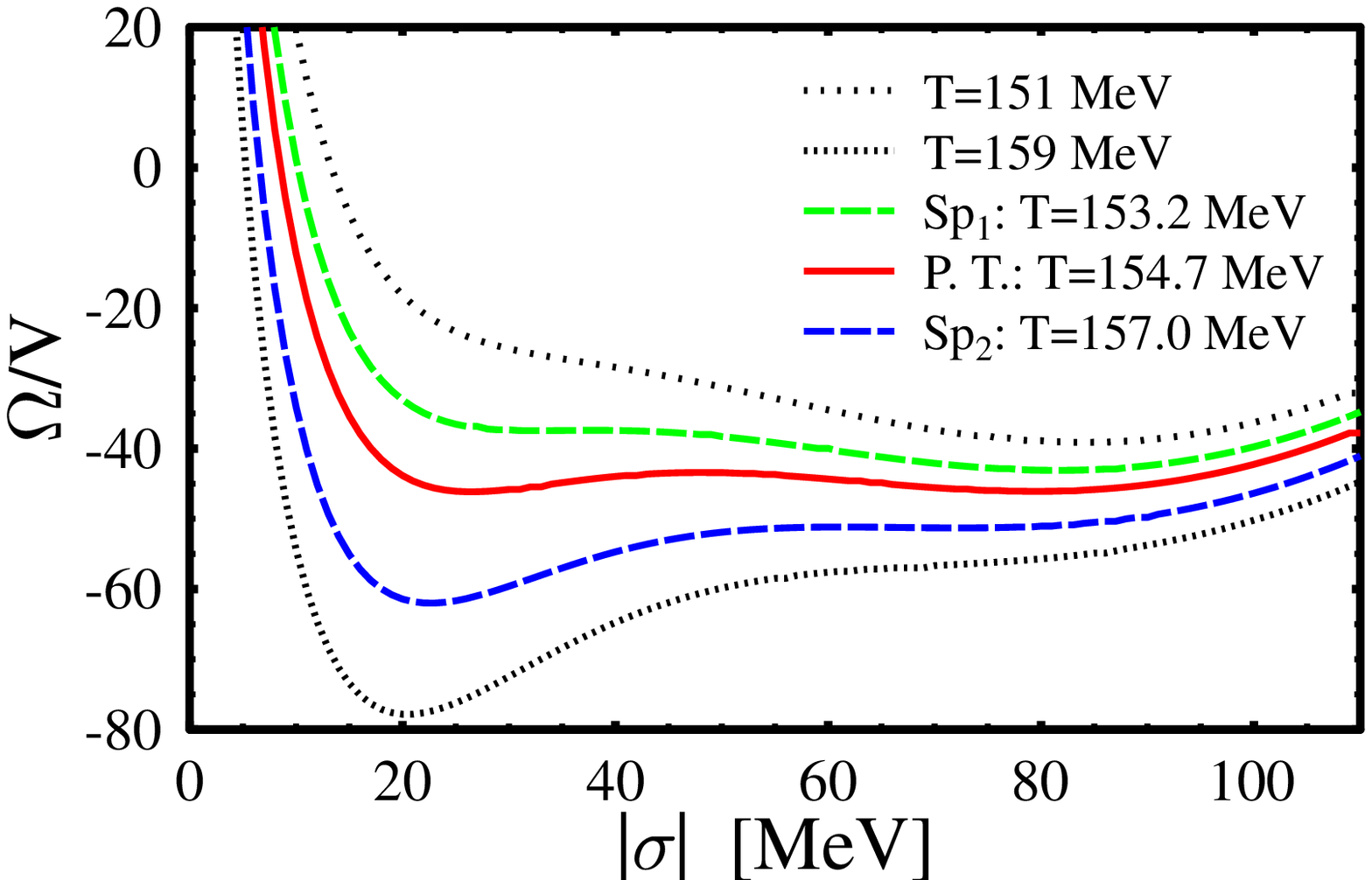}}
\parbox[b]{8.5cm}{
\includegraphics[width=9.2cm,height=6cm]{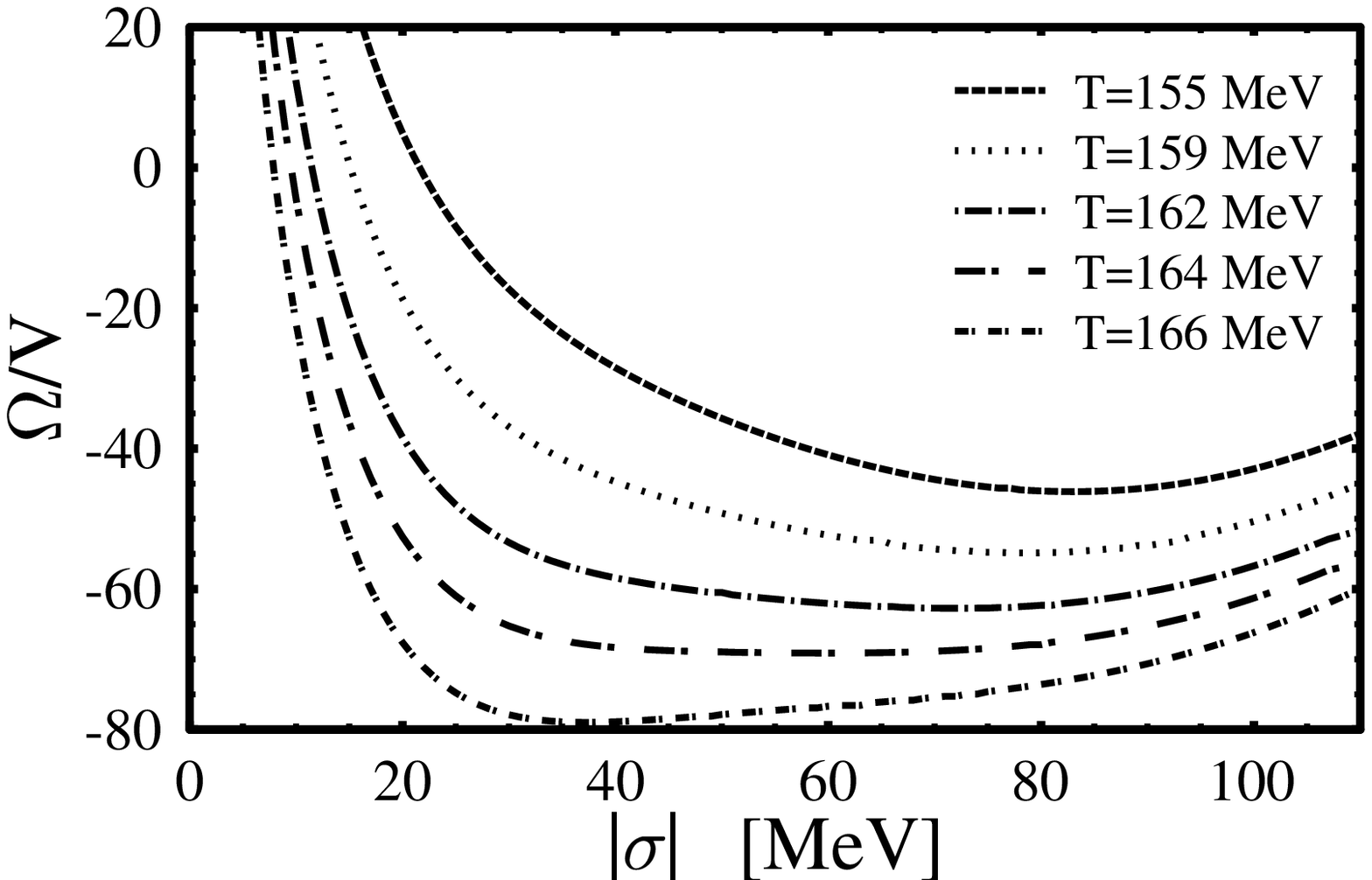}}}
\caption{Effective potential as a 
function of (the order parameter) $\sigma$. %G
The left panel depicts the case with a first-order phase transition for 
$m_{\rm Dec}=0$.
In between both of the spinodals (broken lines) the potential has
two minima  which have the same depth at the phase transition temperature
(full line).
The right panel shows an example for the crossover case, here 
$m_{\rm Dec}=300$ MeV.
The minimum of the potential achieves values of the 
$\sigma$-field of $\approx 20$ MeV
yet at notably higher temperatures than in the first-order phase
transition case.
% for $m_{\rm Dec}=0$ (left) and $m_{\rm Dec}=300$ MeV (right). 
\label{effpot}}
\end{center}
\vspace{-0.5cm}
\end{figure}

For given $T$- and $\mu_q$-values the thermodynamic
potential is given by Eq.\ (\ref{thermpot}).
Minima in the potential characterize the different phases present.
In addition, the knowledge of the potential as a function of the fields, 
i.e.\ away from the minimum, is also of interest, e.g.\ for 
non-equlibrium dynamics of the phase transition \cite{Paech03}. 
In Fig.\ \ref{effpot} the effective potential for two different values of the 
scalar coupling of the baryon resonances is depicted around $T_c$. 
For $m_{\rm Dec}=0$ (left) two distinct minima appear, showing the same
depth at $T_c$
(full line). %G.
Thus, at this temperature the stable phase of the system
jumps from large to low $\sigma$ values. %changes. 
%In contrast, for $m_{\rm Dec}= 300$ MeV
For $m_{\rm Dec}= 300$ MeV, in contrast, %G
there exists only one minimum in the effective potential which 
continuously changes with temperature. 
This characterizes the crossover transition.
%%%%%%%%%%%%%%%%%%%%%%%% potential at mu=0

\begin{figure}[h]
%\vspace*{-3cm}
%\epsfysize=0.8\textheight
\begin{center}
\centerline{\parbox[b]{16.5cm}{
\includegraphics[height=7cm]{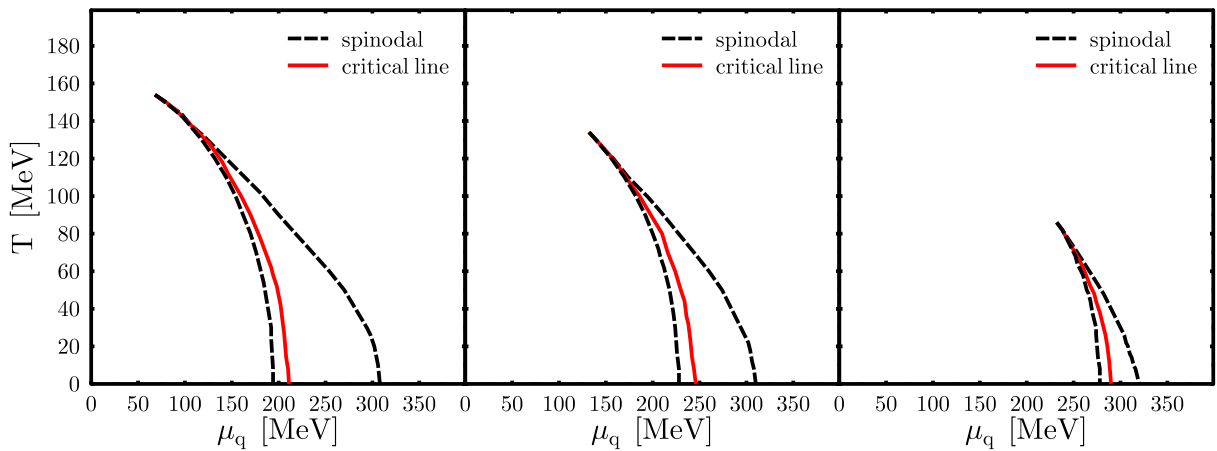}}
}
\vspace{-0.5cm}
\caption{\label{phase_m300}{Phase diagram for $m_{\rm Dec}=300$ MeV and 
increasing %different
values of the decuplet vector coupling 
from left to right ($r_{\rm v}=0, 0.2, 0.4$). %G.
}}
\end{center}
\vspace{-0.5cm}
\end{figure}

We now turn to finite chemical potentials to investigate 
the phase diagram in the whole $T$-$\mu$ plane. Recent lattice QCD
data predict a crossover at vanishing chemical potential 
\cite{Fodor04,Karsch04,deForcrand02} and a critical 
end %G.
point at $T_c \approx 150-170$ MeV and $\mu_{q,c} \approx 100-250$ MeV.
Since several sources of uncertainty remain in these calculations, 
as e.g.\ the large $\pi$ mass or the finite lattice spacings, 
we will also discuss the case of a first-order
phase transition at $\mu=0$.
First, however, %However, first 
we want to consider 
the case with %G.
a crossover at $\mu=0$.
As discussed above this is obtained for $m_{\rm Dec} \geq 260$
MeV. Then, the highest $T_c$ is obtained for a minimal vector 
coupling $r_{\rm v}=0$. We find $T_c \approx 155$ MeV and $\mu_{q,c} \approx 70$ MeV. 
The resulting phase diagram 
is depicted in Fig.\ \ref{phase_m300} (left), 
showing a critical point in the same region as predicted by lattice
QCD. 
However, the critical temperature 
drops %decreases 
much faster in 
our model calculation than in the lattice results. 
Increasing the vector 
coupling of the baryon resonances decreases $T_c$ and increases
$\mu_{q,c}$ (see 
Fig.\ \ref{phase_m300} middle and right). Very
similar behavior is observed when decreasing the scalar coupling 
and keeping the vector coupling constant.
However, for all choices shown in 
Fig.\ \ref{phase_m300} ($r_{\rm v}=0, 0.2, 0.4$), %at $T=0$ 
the chiral phase transition occurs at $\mu_q < 300$ MeV
for $T=0$, which implies %G
a stable chirally restored phase.
Hence, the current form of the model it is not able to obtain 
simultaneously a critical temperature 
in the region predicted by lattice QCD
and a successful description of nuclear
matter properties (density, binding energy).
% Phase digram
%%%%%%%%%%%%%%%%%%%%%%%%% mdek= 300

\begin{figure}[h]
\begin{center}
\centerline{\parbox[b]{8.5cm}{
\includegraphics[width=9.2cm,height=8cm]{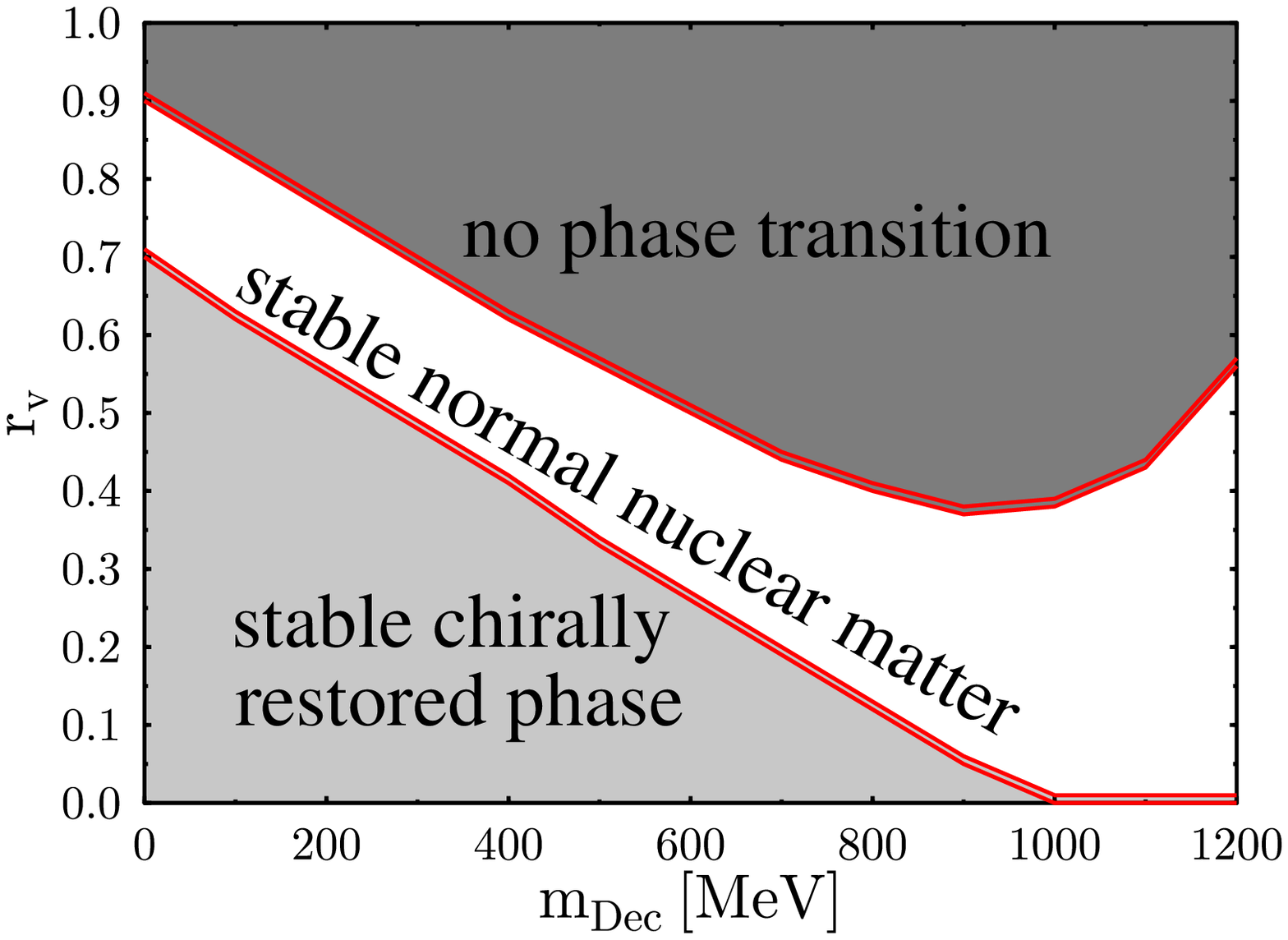}}
\parbox[b]{8.5cm}{
\includegraphics[height=7.0cm]{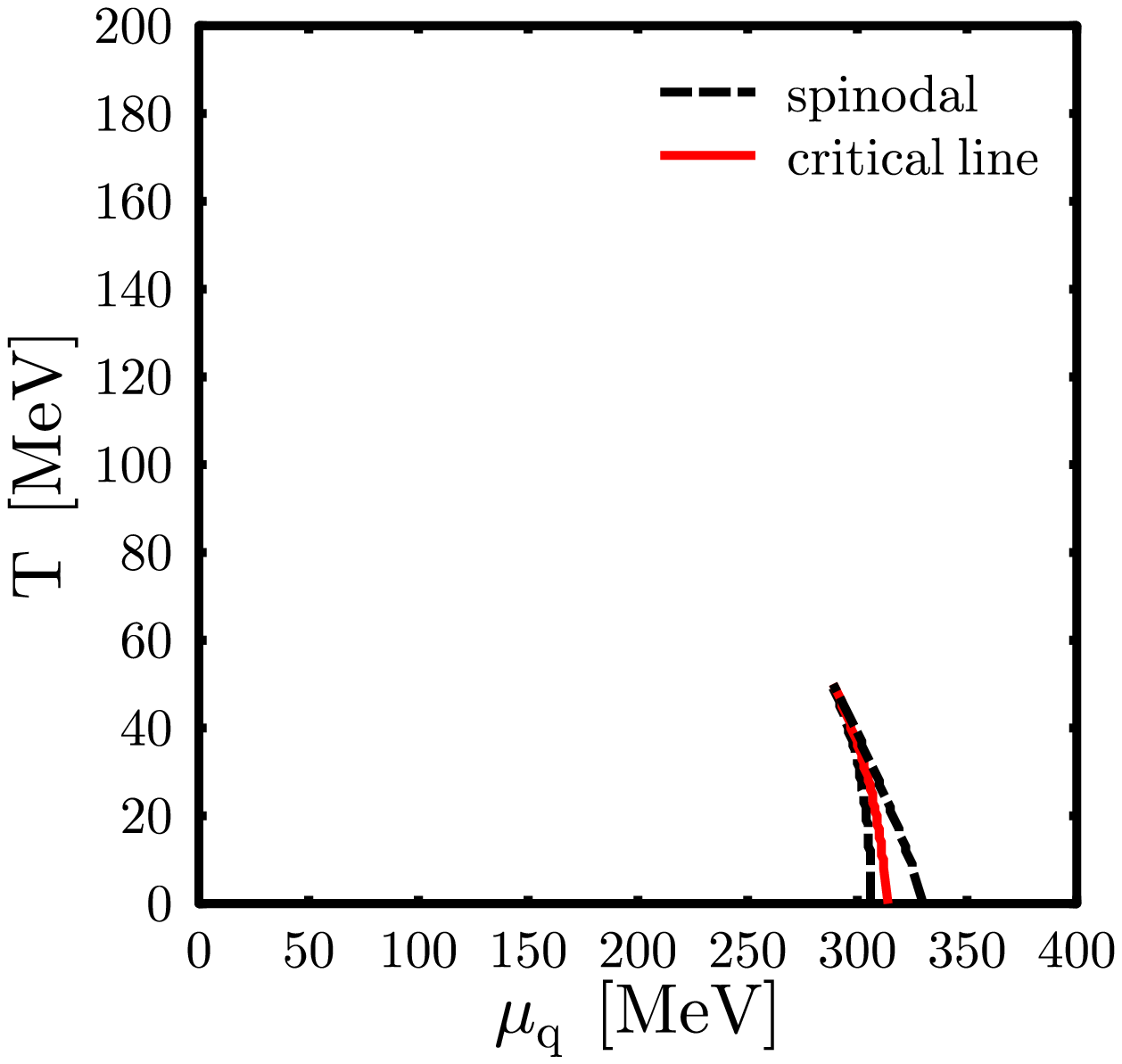}}}
\vspace*{-0.5cm}
\caption{\label{rvmdek_symm}{
(Left) Regions in the $r_{\rm v}$--$m_{\rm Dec}$ parameter space giving  
a first-order phase transition 
with %and
an absolutely stable chirally restored phase,
a chiral phase transition but stable normal nuclear matter 
and a crossover, respectively,
for $T=0$. %G.
(Right) Phase diagram for  
$r_{\rm v}$ and $m_{\rm Dek}$ chosen such that nuclear matter is stable
($m_{\rm Dek}=300$ MeV and $r_{\rm v}=0.5$).}} %G.
\end{center}
\vspace{-0.5cm}
\end{figure}

%
% T=0
%
In Fig.\ \ref{rvmdek_symm} (left) we show the parameter regions,
which, at $T=0$, either 
give a first-order phase transition and an absolutely stable chirally
restored phase, or a chiral phase transition but stable normal nuclear matter,
or no phase transition at all (crossover), respectively.
We observe that the
smaller the explicit symmetry breaking term $m_{\rm Dec}$, i.e.\ the larger 
the scalar coupling of the baryon resonances, %a sufficiently 
the larger the vector coupling constant 
must be chosen to guarantee stable normal nuclear matter.
This is in agreement with the results obtained in 
\cite{Wald87,koso98,Zschiesche:2000ew}.
Insisting that normal nuclear matter 
be stable reduces the maximum critical temperature 
to $T_c \approx 50$ MeV, with a critical chemical potential 
$\mu_{q,c} \approx 290$ MeV. 
This %The 
phase diagram is shown in Fig.\ %figure 
\ref{rvmdek_symm} (right) for $m_{\rm Dec}=300$ MeV and 
$r_{\rm v}=0.5$, the correspondingly lowest possible 
value for $r_{\rm v}$, which  
leads to stable normal nuclear matter. 
The resulting phase diagrams 
for higher values of the explicit symmetry
breaking $m_{\rm Dec}$ and the corresponding smallest possible vector
couplings, look nearly identical. Thus, in the present form 
the chiral model yields $T_c \stackrel {<}{\sim} 50$ MeV, if we insist 
on normal nuclear matter being stable.

\begin{figure}[h]
\vspace*{-0.8cm}
\begin{center}
\centerline{\parbox[b]{8.5cm}{
\includegraphics[width=8.5cm]{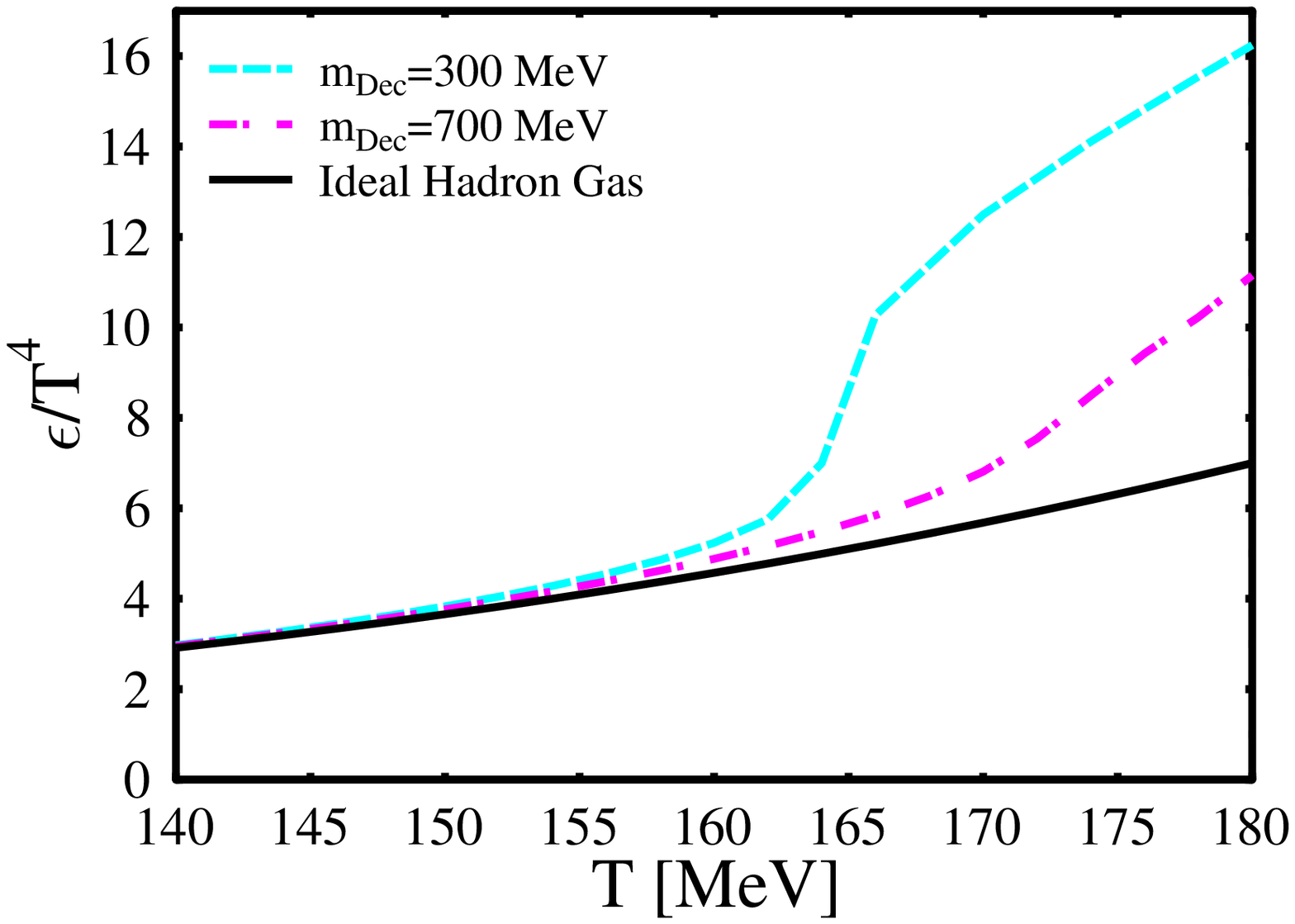}}
\parbox[b]{8.5cm}{
\includegraphics[width=8.5cm]{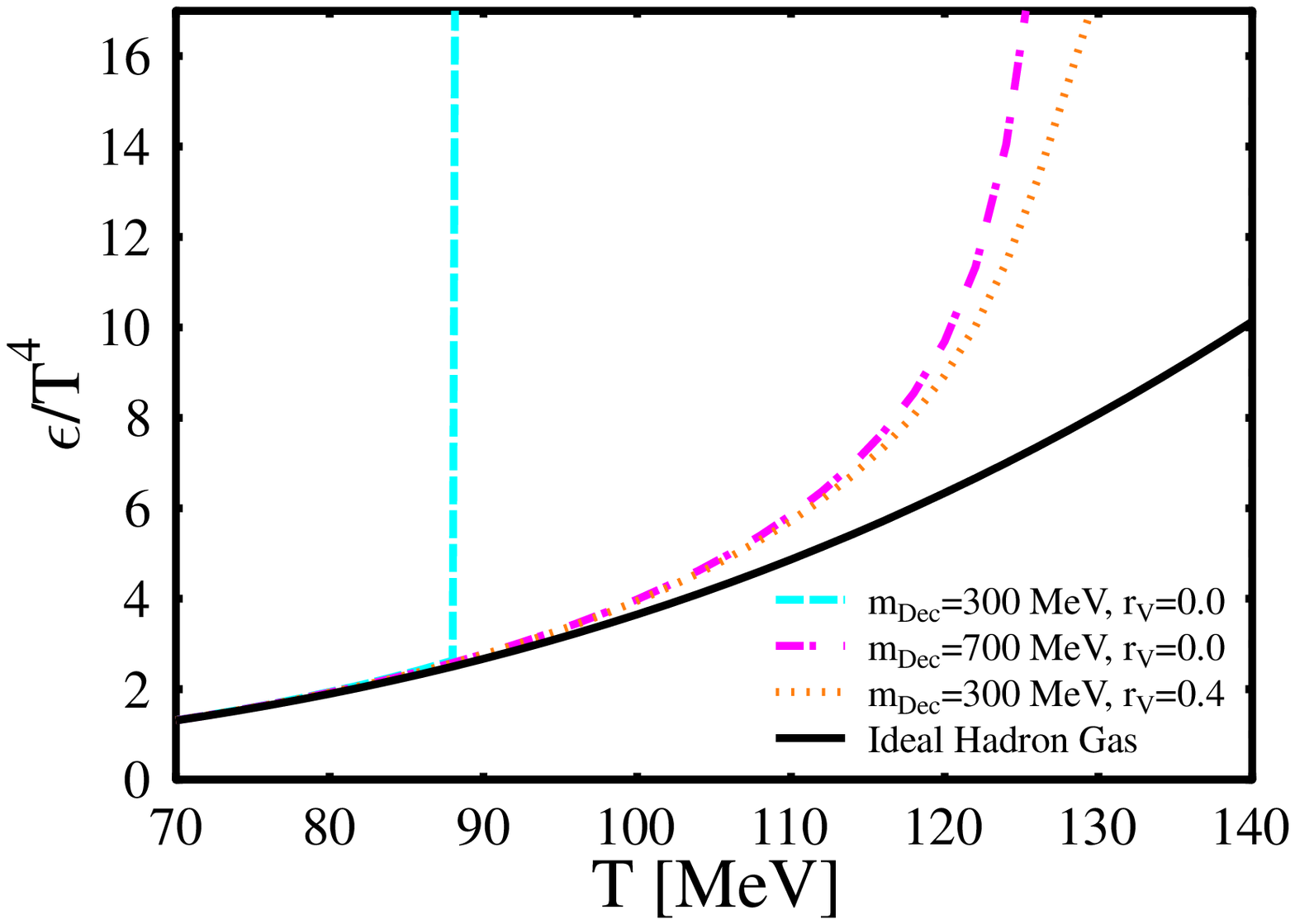}}}
\vspace{-0.5cm}
\caption{\label{endens}{
Energy density over $T^4$ as a function of $T$ with $\mu_q=0$ (left)
and $\mu_q=170$ MeV (right) for different values of $r_{\rm v}$ 
(for $\mu_q \neq 0$) and $m_{\rm Dec}$.
In each figure, one can clearly see the rapid departure
of the model results %G
from the ideal hadron gas curve (full line) in the respective phase
transition region.
}}
\end{center}
\vspace{-0.5cm}
\end{figure}

Examples for the energy density are shown in Fig.\ \ref{endens}
as a function of temperature at constant chemical potential 
and for different values of $r_{\rm v}$ and $m_{\rm Dec}$ (broken lines) as
well as for an ideal hadron gas (full lines).
The left panel shows the case for $\mu_q=0$, where $r_{\rm v}$ 
has %values have
no influence due to the vanishing net baryon density.
The right panel shows the case for $\mu_q=170$ MeV,
illustrating the qualitatively similar effect of increasing either $r_{\rm v}$
or $m_{\rm Dec}$.
For both chemical potentials the interacting and the free gas show
similar properties at low $T$, but exhibit large differences at
higher $T$.
For a first-order phase transition the energy density jumps
to extremely high values (dashed line on right panel).

\begin{figure}[h]
%\vspace*{-3cm}
%\epsfysize=0.8\textheight
\begin{center}
\centerline{\parbox[b]{16.5cm}{
\includegraphics[height=7cm]{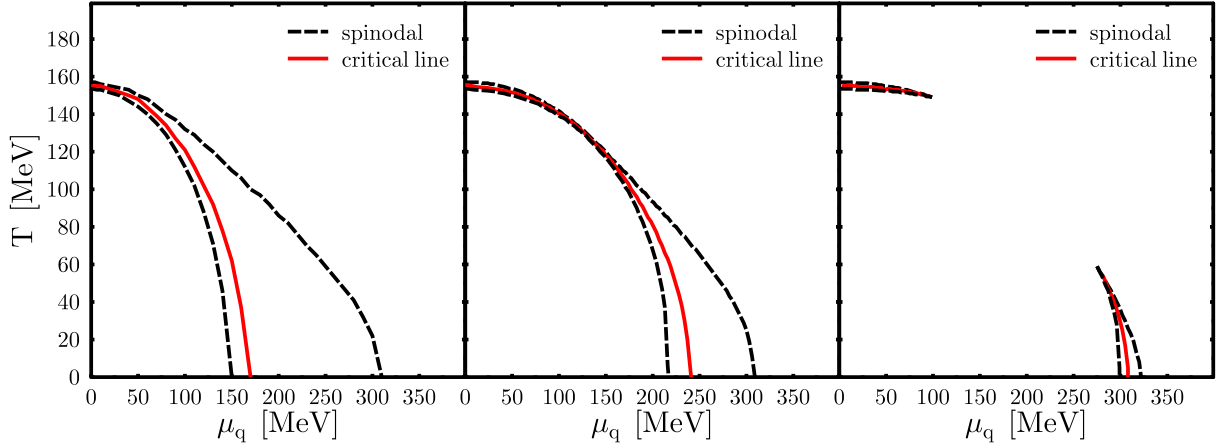}}
}
\vspace{-0.5cm}
\caption{\label{phase_m0}{Phase diagram for 
$m_{\rm Dec}=0$ and $r_{\rm v}=0,0.4, 0.7$ (from left to right). 
For $r_{\rm v}=0.4$ the phase transition weakens at intermediate chemical
potentials and  for $r_{\rm v} \geq 0.5$ 
the phase transition line is disconnected, as can be seen in the
right figure for $r_{\rm v}=0.7$, which 
yields %gives 
stable nuclear matter. 
}}
\end{center}
\vspace{-0.5cm}
\end{figure}

Recent lattice QCD results 
show %predict 
a crossover at $\mu=0$. 
However, as stated above, several uncertainties remain. 
Thus, we also want to consider  
a first-order phase transition 
at $\mu=0$. Then, new and different structures can be observed 
as shown in Fig.\ \ref{phase_m0} for $m_{\rm Dec}=0$.
The line representing the first-order phase transition starts 
from the $T$-axis. For small $r_{\rm v}$ this line continues
down %G.
to the $\mu_q$-axis. However, the phase transition weakens at  
moderate chemical potential, then becomes stronger again 
as $T \rightarrow 0$.  
For $r_{\rm v} \stackrel {>}{\sim} 0.5$ the 
first-order phase transition line ends at $\mu_q \approx 100$ MeV 
and re-appears again at high chemical potentials if  
$r_{\rm v} \leq 0.9$ (cf.\ Fig.\ \ref{rvmdek_symm} left).

\begin{figure}[h]
\vspace*{-0.8cm}
\centerline{
\includegraphics[height=6cm]{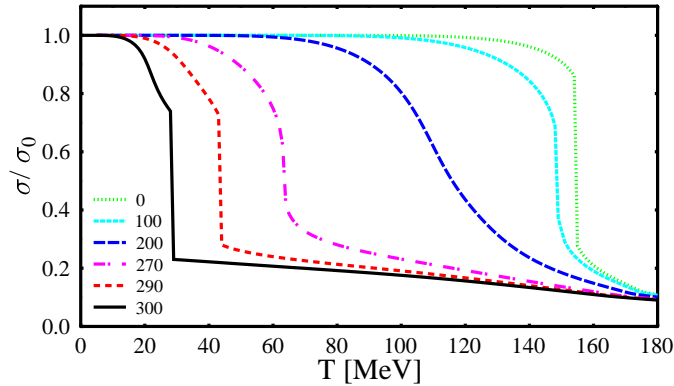}}
\vspace*{-0.5cm}
\caption{\label{sigma_m0_rv07}{Non-strange condensate $\sigma$ 
as a function of temperature for $m_{\rm Dec}=0$, $r_{\rm v}=0.7$ and for
different values of the chemical potential 
$\mu_q=0 \dots 300$ MeV. %G.
%One can clearly see the ''jumps'' oin
For low and high chemical potential the $\sigma$-field
%''
jumps, %''
which corresponds to a first-order phase transition.
%In contrast, 
For intermediate values of $\mu_q$,
instead, %G.
a continuous behavior results.
}}
\end{figure}

The right panel shows the prediction for an acceptable
% simultaneous
description of nuclear matter groundstate properties.
The first-order phase transition line starting from $\mu=0$ ends in a 
critical point $T_c \approx 150$ MeV and $\mu_{q,c} \approx 100$ MeV. 
For 
%higher 
150 MeV $\leq \mu_q \leq 280$ MeV a crossover occurs.
Then a first-order phase transition line appears again, 
reaching down to the $T=0$ axis. This phase diagram 
differs markedly from the lattice results. Figure  
\ref{sigma_m0_rv07} shows how this behavior 
is reflected in the non-strange condensate. 
One can clearly see the %''
jumps %''
in the $\sigma$-field
for low and high chemical potentials and the continuous behavior for
intermediate $\mu_q$.
The phase transition at high temperatures and small chemical potentials
is driven by the abundant production of particle-antiparticle pairs,
which decreases the chiral condensates and thus the effective 
baryon masses. 
Therefrom a further increase of the pair production results.
This is a similar mechanism as the one described in \cite{Thei83}.
At small temperatures and 
high chemical potentials the phase transition is driven by
the rapid increase of the vector density of the baryon resonances,
especially the $\Delta$s, leading to a second minimum in the energy
per particle.

% energy density -> effective deconfinement

\section{IV. Conclusion}
Using a chiral $SU(3)$ model we investigated the dependence of the 
QCD phase diagram on the scalar- and vector-coupling 
of the baryon decuplet. We found that qualitative agreement with
recent lattice results can be obtained, as for example  
a critical point around $T_c \approx 150$ MeV and finite $\mu_{q,c}$.
However, the %very
slow decrease of the phase transition temperature
with increasing chemical potential could not be reproduced. 
Moreover, demanding existence of 
a normal nuclear matter ground-state at $T=0$ reduces $T_c$ 
significantly to approximately  50 MeV. 
Allowing for a first-order phase transition at $\mu=0$,
which cannot unambigiously be excluded from 
current %G
lattice data, 
a rich phase structure is possible, depending on the vector coupling 
of the baryon resonances. If these are chosen to be small then 
a continuous first-order phase transition line from the 
$T$- to the 
$\mu$-axis results.
However, if the vector coupling is increased, 
the first-order phase transition line starting at the $T$-axis 
ends at intermediate chemical potentials. 
Depending on the adopted vector coupling,
a first-order phase transition line re-appears at higher chemical
potentials and lower temperature. Hence, there are two critical
endpoints appearing in the QCD phase diagram
for such a choice of parameters. %G

These results show that the simultaneous description of 
the phase diagram and nuclear matter saturation gives strong
constraints on the model. Even more constraints may appear if 
also neutron stars are considered \cite{zschiesche04}.
Since many different models 
with different equations of state give a good description of nuclear
matter saturation and neutron stars, such additional constraints 
are urgently needed -- although there is still considerable uncertainty 
in the lattice
calculations. %results.
However, further improvement can be expected
in the near future in this area and thus more reliable results will
appear. Then one should be able to find connections between the
phase diagram and the structure of specific models.

Our results also show that a very diverse structure of the phase
diagram is possible. To get a better understanding of 
the characteristics of the phase transition at high 
chemical potential, it would be very helpful to gain knowledge 
on the latent heat and correlation lengths 
as a function of chemical potential from 
lattice calculations.
Then one might be able to pin down the phase diagram in a more 
quantitative fashion. 

Although our study neglected many contributions, especially the 
pions and the higher resonances, some very interesting conclusions
appear.
First of all, the baryon resonances may very well drive the 
chiral phase transition and give a structure in agreement with 
lattice QCD. The question how the 
QCD phase diagram with realistic pion masses looks like is still open.
As was shown in \cite{Stephanov04},
several different approaches, which do not take into account baryon
resonances, yield much lower critical temperatures.
This clearly suggests an important role of baryonic degrees of freedom 
for the characteristics of the chiral phase transition. 
The stronger curvature of $T_c(\mu)$ as compared to
lattice results might be due to only  
including the baryon decuplet.
%Furthermore, due to the 'explosion' of density and energy density 
%at the first-order chiral phase transition, the validity of an hadronic
%description of the thermodynamic state becomes quite 
%questionable simultaneously.
At the same time, however, the validity of a hadronic
description of the thermodynamic state becomes quite 
questionable due to the 'explosion' of density and energy density 
at the first-order chiral phase transition. %G

So far we only took into account the contribution of
the resonances by coupling the baryon decuplet to the fields. 
The extension to more degrees of freedom is in progress.
It will be very interesting to see how the results obtained here may be changed
by a larger 
hadronic %G.
spectrum. 

Another very promising investigation is the comparison to the 
equation of state and the chiral condensate 
as obtained from lattice calculations with a distorted mass 
spectrum -- as proposed in \cite{KarschTawfika,KarschTawfikb}. 
There it was shown that a non-interacting resonance gas can give a 
good description of the lattice equation of state.
Since close to $T_c$ the effect of interactions should definitely 
be present, however, the comparisons to the lattice results 
should help to disentangle the field contributions from those  
of an ideal gas and to learn more about the nature 
of the interactions close to the phase boundary. 
In particular, within the chiral $SU(3)$ model one is able to study 
the behavior of the chiral condensate and thermodynamic quantities 
simultaneously. 
In addition, it is possible to make contact 
between the theory of the phase diagram and experimental 
observables,  
e.g.\ %for example 
by determining in one model the 
chemical freeze-out temperature in relativistic heavy
ion collisions -- as obtained from fits to the measured particle
ratios  -- and the phase 
transition temperature \cite{Zsch02,Zschiesche2004}. 

\begin{acknowledgements}
%%%%%%%%%%%%%%%%%%%%%%%%%%%%%%%%%%%%%%%%%%%%%%%%%%%%%%%%%%%%%%%%%%%
The authors are grateful to Adrian Dumitru, Carsten Greiner, J\"urgen 
Schaffner-Bielich, and Kristof Redlich for 
fruitful discussions. This work is supported by Deutsche
Forschungsgemeinschaft (DFG), Gesellschaft f\"ur Schwerionenforschung
(GSI) and Bundesministerium f\"ur Bildung und Forschung (BMBF).
This work used computational resources provided by the Center for Scientific 
Computing (CSC) at the University of Frankfurt, Germany.
\end{acknowledgements}

%%%%%%%%%%%%%%%%%%%%%%%%%%%%%%%%%%%%%%%%%%%%%%%%%%%%%%%%%%%%%%%%%%%
%\appendix
%\section{}
%\label{append}
%%%%%%%%%%%%%%%%%%%%%%%%%%%%%%%%%%%%%%%%%%%%%%%%%%%%%%%%%%%%%%%%%%%
\bibliography{chiral}
\bibliographystyle{h-elsevier2.bst}
\end{document}